\documentclass[twocolumn,showkeys,amsmath,amssymb,pra]{revtex4}

\usepackage{graphicx,amssymb}
\usepackage{dcolumn}
\usepackage{bm}

\newcommand{\beq}{\begin{equation}}
\newcommand{\eeq}{\end{equation}}
\newcommand{\bea}{\begin{eqnarray}}
\newcommand{\eea}{\end{eqnarray}}

\begin{document}

\title{Precession of Vortices in Dilute Bose-Einstein Condensates at Finite
Temperature}

\author{B. G. Wild and D. A. W. Hutchinson}
\affiliation{The Jack Dodd Centre for Quantum Technology, Department of Physics, University of Otago, Dunedin 9016, New Zealand}

\begin{abstract}
We demonstrate that the precessional frequencies
of vortices in Bose Einstein condensates (BECs) are determined by
a conservation law, and not by the lowest lying excitation energy mode. We determine the precessional frequency for a single off-axis vortex and vortex lattices in BECs using the continuity equation, and solve this self-consistently with the time-independent Hartree-Fock-Bogoliubov (HFB) equations in the rotating frame. We find agreement with zero temperature calculations (Bogoliubov approximation), and a smooth variation in the precession frequency as the temperature
is increased. Time-dependent solutions confirm the validity of these predictions.
\end{abstract}

\maketitle

Quantised vortices are perhaps the single most striking manifestation
of superfluidity. The dynamics of vortices in Bose Einstein condensates (BECs) have been of particular interest \cite{Experimental} with earlier theoretical work \cite{Theoretical,Dodd,Feder,Svidinsky} concentrating on zero temperature,
neglecting the effects of the non-condensate on the overall
dynamics of the vortex. The Bogoliubov spectrum in this case has a lowest-lying
energy that is negative, the anomalous mode \cite{Dodd,Feder,Svidinsky}.
In the frame rotating at the frequency corresponding to this energy, the lowest-lying energy in the Bogoliubov excitation spectrum is zero, leading to the conclusion that this mode frequency corresponds to the precessional frequency of
the vortex \cite{Dodd,Feder,Svidinsky}.

To go to finite temperature, one can introduce the Hartree-Fock-Bogoliubov (HFB) treatment \cite{Griffin}. Solving the HFB equations in the laboratory frame for an on-axis vortex at finite temperature in the so-called Popov approximation \cite{Griffin} yields a positive lowest lying energy mode, referred to as the lowest core localised state (LCLS)\cite{Isoshima,Virtanen1}. This treatment was generalised to an off-axis vortex \cite{Virtanen2} and provided a lower bound for the LCLS energy. In both cases it is argued that the thermal cloud acts as an effective potential, thereby stabilising the vortex, but both fail to take into account the dynamics of the thermal cloud itself. The association of the LCLS energy with the precessional frequency of the vortex \cite{Virtanen1} led to the conclusion that the vortex precesses in the direction opposite to the condensate flow around the core. This is inconsistent with experiment \cite{Feder2} and also seems intuitively unreasonable, since zero temperature predictions would suggest otherwise \cite{Dodd,Feder,Svidinsky} and one might reasonably expect continuous variation with temperature. However, in later work on off-axis vortices \cite{Isoshima2}, Isoshima \emph{et. al.} conclude that the precessional frequencies and LCLS energies are indeed uncorrelated, but assert rather that the LCLS energy is responsible for the inward/outward motion of vortices. They have no explicit equation for the prediction of precessional frequencies, but rely on the assertion that the LCLS energy depends linearly on the rotational frequency, and solve the problem by iteration.

Here we make use of the continuity equation to make an a priori prediction of the vortex precessional frequency and solve the HFB equations self-consistently in the frame rotating at this frequency. This equation is also valid for vortex lattices, and stationary solutions can be found in the frame rotating at this precessional frequency, provided the vortex interactions are not too strong.

In a frame rotating with angular frequency $\Omega$, the time-dependent HFB formalism \cite{Buljan} consists of the generalised Gross-Pitaevskii equation (GGPE) for our condensate wavefunction $\Phi$
\begin{equation}
\begin{array}{lcl}
i\hbar\frac{\partial}{\partial t}\Phi(\mathbf{r},t) & = & \left(\hat{h}_{\Omega}(\mathbf{r})+g\left(n_{c}(\mathbf{r},t)+2\tilde{n}(\mathbf{r},t)\right)\right)\Phi(\mathbf{r},t),\\ & & 
+g\tilde{m}(\mathbf{r},t)\Phi^{*}(\mathbf{r},t)\end{array}\label{eq:1}
\end{equation}
and the Bogoliubov-de Gennes equations (BdGEs)
\begin{equation}
i\hbar\frac{\partial}{\partial t}\left[\begin{array}{c}
u_{q}(\mathbf{r},t)\\
v_{q}(\mathbf{r},t)\end{array}\right]=\left[\begin{array}{cc}
\mathcal{\hat{L}}_{\Omega}(\mathbf{r},t) & \mathcal{\hat{M}}(\mathbf{r},t)\\
-\mathcal{\hat{M}}^{*}(\mathbf{r},t) & -\mathcal{\hat{L}}_{\Omega}^{*}(\mathbf{r},t)\end{array}\right]\left[\begin{array}{c}
u_{q}(\mathbf{r},t)\\
v_{q}(\mathbf{r},t)\end{array}\right],\label{eq:2}\end{equation}
with
\begin{equation}
\begin{array}{lcl}
\hat{h}_{\Omega}(\mathbf{r}) & = & -\frac{\hbar^{2}}{2m}\nabla^{^{2}}+i\hbar \mathbf{\Omega}.(\mathbf{r}\times\mathbf{\nabla})+V_{T}(r),\\
\mathcal{\hat{L}}(\mathbf{r},t) & = & \hat{h}(\mathbf{r})+2g\left(n_{c}(\mathbf{r},t)+\tilde{n}(\mathbf{r},t)\right)-\mu,\\
\mathcal{\hat{M}}(\mathbf{r},t) & = & g\left(\Phi^{^{2}}(\mathbf{r},t)+\tilde{m}(\mathbf{r},t)\right).\end{array}\label{eq:3}\end{equation}
Here $g=4\pi\hbar^{2}a_{s}/m$ is the interaction strength, $a_{s}$ the s-wave scattering length, $m$ the atomic mass. The condensate density $n_{c}$, thermal density $\tilde{n}$ and anomalous density $\tilde{m}$
have their usual definition \cite{Griffin}.

Defining the current density $\mathbf{j}(\mathbf{r},t)\equiv-i\frac{\hbar}{2}\left(\Phi^{*}(\mathbf{r},t)\mathbf{\nabla}\Phi(\mathbf{r},t)-\Phi(\mathbf{r},t)\mathbf{\nabla}\Phi^{*}(\mathbf{r},t)\right)$,
one obtains from (\ref{eq:1}) the continuity equation \begin{equation}
\begin{array}{c}
\hbar\frac{\partial}{\partial t}\left|\Phi(\mathbf{r},t)\right|^{2}+\frac{\hbar}{m}\mathbf{\nabla}.\mathbf{j}(\mathbf{r},t)=\hbar\mathbf{\Omega}.(\mathbf{r}\times\mathbf{\nabla})\left|\Phi(\mathbf{r},t)\right|^{2}\\
-ig\left(\tilde{m}(\mathbf{r},t)\Phi^{*^{2}}(\mathbf{r},t)-\tilde{m}^{*}(\mathbf{r},t)\Phi^{2}(\mathbf{r},t)\right).\end{array}\label{eq:8}\end{equation}
The corresponding time-independent HFB equations are given by \cite{Griffin}
\begin{equation}
\mu\Phi(\mathbf{r})=\left(\hat{h}_{\Omega}(\mathbf{r})+g\left(n_{c}(\mathbf{r})+2\tilde{n}(\mathbf{r})\right)\right)\Phi(\mathbf{r})+g\tilde{m}(\mathbf{r})\Phi^{*}(\mathbf{r}),\label{eq:9}\end{equation}
\begin{equation}
\epsilon_{q}\left[\begin{array}{c}
u_{q}(\mathbf{r})\\
v_{q}(\mathbf{r})\end{array}\right]=\left[\begin{array}{cc}
\mathcal{\hat{L}}_{\Omega}(\mathbf{r}) & \mathcal{\hat{M}}(\mathbf{r})\\
-\mathcal{\hat{M}}^{*}(\mathbf{r}) & -\mathcal{\hat{L}}_{\Omega}^{*}(\mathbf{r})\end{array}\right]\left[\begin{array}{c}
u_{q}(\mathbf{r})\\
v_{q}(\mathbf{r})\end{array}\right],\label{eq:10}\end{equation}
where $\mu$ is the chemical potential, and $\epsilon_{q}$ are the
quasi-particle energies.

We note, however, that the quasi-particle amplitudes
calculated using this formalism are not, in general, orthogonal to the condensate \cite{Morgan}. This
can cause anomalous predictions for $\Omega$ in various temperature
regimes. We can derive a suitable orthogonal formalism, i.e. one where $\int d\mathbf{r}\Phi^{*}(\mathbf{r},t)u_{q}(\mathbf{r},t)=\int d\mathbf{r}\Phi(\mathbf{r},t)v_{q}(\mathbf{r},t)=0$
as follows: We write the field operator \cite{Castin} as $\hat{\psi}(\mathbf{r},t)=\hat{a}_{c}(t)\phi(\mathbf{r},t)+\hat{\eta}(\mathbf{r},t)$,
where $\hat{a}_{c}(t)=\int dr\hat{\psi}(\mathbf{r},t)\phi^{*}(\mathbf{r},t)$ and $\hat{a}_{c}^{\dag}(t)$
are the annihilation/creation operators corresponding to the condensate. $\phi(\mathbf{r},t)$ is the condensate wavefunction normalised to unity such that $\Phi = \sqrt{N_{c}} \phi$ where $N_{c}$ is the number of condensate atoms, and $\hat{\eta}(\mathbf{r},t)=\int dr^{\prime}Q(\mathbf{r},\mathbf{r}^{\prime})\hat{\psi}(\mathbf{r}^{\prime},t)$ is the fluctuation operator corresponding to the thermal cloud, with 
$Q(\mathbf{r},\mathbf{r}^{\prime})\equiv\delta(\mathbf{r},\mathbf{r}^{\prime})-\phi(\mathbf{r},t)\phi^{*}(\mathbf{r}^{\prime},t)$. Using the commutation relations $\left[\hat{a}_{c}(t),\hat{a}_{c}^{\dag}(t)\right]=1$,
$\left[\hat{\eta}(\mathbf{r},t),\hat{\eta}^{\dag}(\mathbf{r}^{\prime},t)\right]=Q(\mathbf{r},\mathbf{r}^{\prime})$
and $\left[\hat{a}_{c}(t),\hat{\eta}^{\dag}(\mathbf{r},t)\right]=0$, and applying the Bogoliubov transformation and  the usual mean-field approximations\cite{Griffin}, one obtains the orthogonal time-independent HFB equations
consisting of the GGPE (\ref{eq:9}) and the orthogonal
BdGEs given by (\ref{eq:10}), but where the $\mathcal{\hat{L}}_{\Omega}(\mathbf{r})$ and $\mathcal{\hat{M}}(\mathbf{r})$ operators are replaced by
\begin{equation}
\begin{array}{lcl}
\mathcal{\hat{L}}_{\Omega}(\mathbf{r},\mathbf{r}^{\prime}) & \rightarrow & \int d\mathbf{r}^{\prime}Q(\mathbf{r},\mathbf{r}^{\prime})\mathcal{\hat{L}}_{\Omega}(\mathbf{r}^{\prime})\\
\mathcal{\hat{M}}(\mathbf{r},\mathbf{r}^{\prime}) & \rightarrow & \int d\mathbf{r}^{\prime}Q(\mathbf{r},\mathbf{r}^{\prime})\mathcal{\hat{M}}(\mathbf{r}^{\prime})\end{array}\label{eq:5}\end{equation}
This ensures the orthogonality of the condensate and thermal states. In view of this orthogonality, making the assignment $u_{0}\rightarrow\phi$, $v_{0}\rightarrow-\phi^{*}$ satisfies the BdGEs (\ref{eq:10}) with a zero energy. Thus the lowest excitation energy is zero (which is certainly not the case for ordinary HFB), with quasi-particle amplitudes proportional to the ground state, thereby satisfying Goldstone's theorem. However the effect on higher energy modes is negligible, and the energy gap \cite{Hohenberg} still remains an issue.

All time-independent calculations presented here were performed using this orthogonal HFB. The results without the orthogonalisation are essentially the same, appart from some slightly anomalous behaviour in $\Omega$ for vortices close to the axis at low temperature, where the orthogonality of the solutions becomes important. We note that the Popov approximation, as opposed to HFB, violates particle conservation, and is therefore not a suitable formalism for time-dependent simulations. HFB can also be derived from a variational standpoint, and represent a "conserving" approximation \cite {Griffin}. It is from this basis that we use HFB for our simulations. As a validity check, time dependent simulations where the BEC is initially displaced from the axis yield Kohn mode \cite{Dobson} oscillations at precisely the radial trapping frequency.

To proceed, we write the continuity equation in terms of the real and imaginary parts of the condensate wavefunction. Defining $R(\mathbf{r})\equiv\textrm{Re}(\Phi(\mathbf{r}))$ and $I(\mathbf{r})\equiv\textrm{Im}(\Phi(\mathbf{r}))$,
and noting that in the time-independent case $\frac{\partial}{\partial t}\left|\Phi(\mathbf{r},t)\right|^{2}=0$,
we obtain\begin{equation}
\begin{array}{c}
\mathbf{\Omega}.\left(R(\mathbf{r}\times\mathbf{\nabla})I+I(\mathbf{r}\times\mathbf{\nabla})R\right)=\frac{\hbar}{2m}\left(R\nabla^{2}I-I\nabla^{2}R\right)\\
+i\frac{g}{2\hbar}\left(\tilde{m}\Phi^{*^{2}}-\tilde{m}^{*}\Phi^{2}\right).\end{array}\label{eq:11}\end{equation}

Our model consists of a dilute Bose gas consisting of 2000 $^{87}\textrm{Rb }$atoms in an axially symmetric harmonic trap, with radial and axial harmonic trapping frequencies of $\Omega_{r}=2\pi\times10\textrm{Hz}$ and $\Omega_{z}=2\pi\times400\textrm{Hz}$ respectively. Thus the axial confinement is sufficiently strong that all excited axial states may be neglected. Hence the BEC may be treated as two-dimensional from a computational viewpoint, although the trap geometry is not critical for our conclusions. Thus, restricting ourselves to two dimensions, and defining\[
A(r,\theta)\equiv \left(R\frac{\partial R}{\partial\theta}+I\frac{\partial I}{\partial\theta}\right)\]
and \[
\begin{array}{c}
B(r,\theta)\equiv \frac{\hbar}{2m}\left[r^{2}\left(R\frac{\partial^{2}I}{\partial r^{2}}-I\frac{\partial^{2}R}{\partial r^{2}}\right)+r\left(R\frac{\partial I}{\partial r}-I\frac{\partial R}{\partial r}\right)\right]\\
+\frac{\hbar}{2m}\left(R\frac{\partial^{2}I}{\partial\theta^{2}}-I\frac{\partial^{2}R}{\partial\theta^{2}}\right)+i\frac{gr^{2}}{2\hbar}\left(\tilde{m}\Phi^{*^{2}}-\tilde{m}^{*}\Phi^{2}\right)\end{array}\]
and integrating over all space, we find that\begin{equation}
\Omega=\int_{0}^{2\pi}\int_{0}^{\infty}A(r,\theta)B(r,\theta)rdrd\theta/\int_{0}^{2\pi}\int_{0}^{\infty}r^{2}\left(A(r,\theta)\right)^{2}rdrd\theta.\label{eq:14}\end{equation}

We can now solve equations (\ref{eq:9}), (\ref{eq:10}) and (\ref{eq:14})
self-consistently in the frame rotating at angular frequency $\Omega$
to find stationary solutions for a precessing vortex. A vortex at
position $\left(r_{1},\theta_{1}\right)$ may be created by expanding
the condensate wave function in terms of modified Laguerre basis functions $\left\{ \chi_{ln}^{(1)}(r,\theta)\right\} $
as follows:\begin{equation}
\Phi(r,\theta)=\sum_{ln}^{(1)}a_{ln}\chi_{ln}^{(1)}(r,\theta),\label{eq:19}\end{equation}
provided $\left(r_{1},\theta_{1}\right)$ is not a root of $\xi_{ln}(r,\theta)$,
where we define the computational basis functions \begin{equation}
\chi_{ln}^{(1)}(r,\theta)\equiv\xi_{ln}(r,\theta)-\frac{\xi_{ln}(r_{1},\theta_{1})}{\xi_{10}(r_{1},\theta_{1})}\xi_{10}(r,\theta),\label{eq:20}\end{equation}
and where\begin{equation}
\xi_{ln}(r,\theta)=\frac{e^{il\theta}}{\sqrt{2\pi}}\left(\frac{2n!}{(n+\left|l\right|)!}\right)^{1/2}e^{-r^{2}/2}r^{\left|l\right|}L_{n}^{\left|l\right|}(r^{2}),\label{eq:22}\end{equation}
with $L_{n}^{\left|l\right|}(x)$ a modified Laguerre polynomial of order $n$. The superscript $(1)$ in the summation means
exclude $\left(n,l\right)=\left(0,1\right)$, and indicates a single vortex. We derive (\ref{eq:19}) and (\ref{eq:20}) by expanding the condensate wave function in terms of the complete
Laguerre basis $\left\{ \xi_{ln}(r,\theta),n=0,\ldots,l=0,\pm1,\ldots\right\} $
and considering that $\Phi(r_{1},\theta_{1})=\sum_{ln}a_{ln}\xi_{ln}(r_{1},\theta_{1})=0$.

This procedure may be extended to $N_{v}$ vortices $\left\{ \left(r_{1},\theta_{1}\right),\ldots,\left(r_{N_{v}},\theta_{N_{v}}\right)\right\} $ by
an iterative process, where we write\begin{equation}
\Phi(r,\theta)=\sum_{ln}^{(N_{v})}a_{ln}\chi_{ln}^{(N_{v})}(r,\theta),\label{eq:23}\end{equation}
where we have defined\begin{equation}
\chi_{ln}^{(N_{v})}(r,\theta)\equiv\chi_{ln}^{(N_{v}-1)}(r,\theta)-\frac{\chi_{ln}^{(N_{v}-1)}(r_{N_{v}},\theta_{N_{v}})}{\chi_{N_{v}0}^{(N_{v}-1)}(r_{N_{v}},\theta_{N_{v}})}\chi_{N_{v}0}^{(N_{v}-1)}(r,\theta).\label{eq:24}\end{equation}

Stationary solutions are found in the single vortex case, and in the
multiple vortex case, provided the interactions between the vortices
are not too strong. 


\begin{figure}
\includegraphics[
  scale=0.3,
  angle=0]{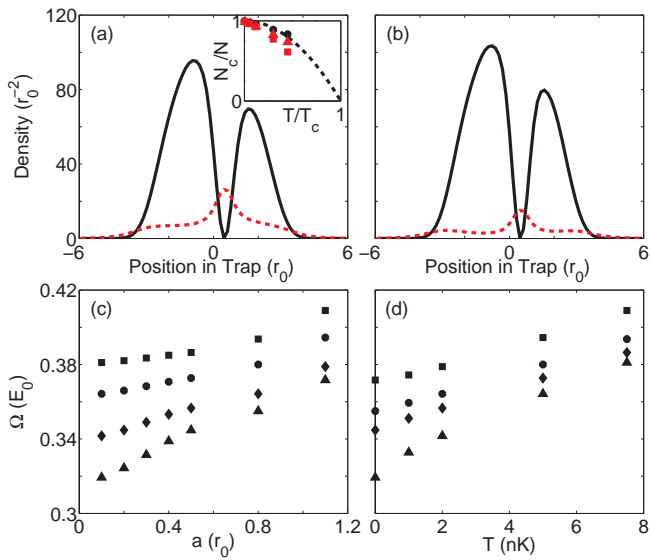}

\caption{(colour online)(a) Popov condensate (solid line) and thermal densities (dashed line) in the plane passing through the vortex, and (insert) condensate fraction for the ideal gas (dashed line), HFB (circles) and Popov calculations for condensate with Bogoliubov mode cutoff of 100 (triangles), and 1639 (squares). (b) HFB condensate (solid line) and thermal densities (dashed line) in the plane passing through the vortex, (c) Precessional frequency $\Omega$ versus vortex position $a$ for HFB at temperatures 0nK, 2nK, 5nK and 7.5nK (triangles, diamonds, circles, squares), (d) Precessional frequency $\Omega$ versus temperature $T$ for HFB at vortex positions 0.1, 0.5, 0.8 and 1.1 (triangles, diamonds, circles, squares). All units are given in radial harmonic oscillator units $r_{0}=\sqrt{\hbar /m\Omega_{r}}$ for distance and $E_{0}=\hbar \Omega_{r}/2$ for energies and frequencies.
\label{cap:fig1}}
\end{figure}

Figures 1(a) and 1(b) show the condensate and non-condensate density
profiles in the plane passing through the vortex for the Popov and HFB
cases respectively for a vortex situated at $a=0.5$ radial harmonic oscillator
units from the axis, at temperature $T_{c}=5\textrm{nK}$. Note in particular
that the thermal density for HFB is considerably less than for Popov. This is due to the upward shift in the lower-lying excitation energies.

One would expect the precessional frequency $\Omega$ to vary smoothly
with temperature $T$, and the time-independent off-axis calculations
for a variety of different vortex positions for the Popov and HFB cases reveal that this is indeed the case, also being
consist with off-axis GPE calculations, which coincide almost
exactly with the HFB $T=0$ calculations (see figures 1(c), 1(d)). This is in stark contrast with previous \cite {Isoshima,Virtanen1,Virtanen2} inferences regarding vortex precession based upon on-axis HFB-Popov calculations with static thermal clouds. These calculations therefore resolve this conflict.

One would also expect the precessional frequency $\Omega$ to vary
smoothly with $a$, so to check the consistency of the on-axis and
off-axis calculations, one needs to determine $\Omega$ for the on-axis
case by extrapolation. Solving for the off-axis case in the frame rotating at the extrapolated
precessional frequency for the Popov approximation, using (\ref{eq:14}), one finds consistency in the LCLS energies. These energies also correspond to the energies calculated perturbatively in \cite{Fetter} with which we find very good agreement at low temperature. One also finds absolutely no correlation between
the precessional frequencies and the LCLS energies in agreement with \cite{Isoshima2}. The excitation due to $\epsilon_{LCLS}$
is given by the mode density corresponding to the LCLS mode, and has
nothing to do with precession. The precessional frequencies are determined
by the conservation of flow (the continuity equation), and hence by
equation (\ref{eq:14}). These calculations also predict precession
in the same sense as the circulation, as one would expect, contrary
to \cite{Virtanen1}. It has been suggested that the vortex precession inferred from previous calculations might be associated with the nutation of the vortex about the artificially static effective pinning potential due to interactions of the condensate with the thermal cloud. To investigate this, time dependent simulations were performed with the thermal density initially rotating with the core at the precessional frequency, but where the rotational frequency of the thermal cloud is gradually reduced. One finds that the vortex core initially slows down with the pinning potential to a certain point, depending on the strength of the pinning potential of thermal density. In principle, given a sufficiently strong pinning potential, one should be able to stop the precession comletely, though to date this has not been achieved. No mutation of the vortex (i.e. precession of the vortex about the pinning potential) has been observed in any of the simulations. We are left to conclude that the LCLS energies play no part in precession or nutation of the vortex and the LCLS corresponds to a collective response of the gas unassociated with the precession of the vortex. We note further that $\epsilon_{LCLS}$ varies continuously with $T$, and we see that $\left.\epsilon_{LCLS}\right|_{T\approx0}$ always takes on some small positive value, associated with the quantum depletion.

We also find consistency in the condensate fractions between the on-axis
and off-axis calculations, and the insert in Figure 1(a) shows a plot of the condensate
fraction $N_{c}$ versus $T(\textrm{nK})$ for the off-axis case for
HFB, and the on-axis case for Popov for a mode cutoff of 100. Additional
points are shown for Popov for mode cutoff of 1639. Shown also is $N_{c}$
versus $T$ for the ideal gas with $T_{c}\approx16.7\textrm{nK}$ . The results reveal
that the mode cutoff of 100 is sufficient for $T\lesssim7.5\textrm{nk}$,
but the results for $T\gtrsim T_{c}/2$ become unreliable. At these temperatures the HFB formalism is unreliable in this regime anyway \cite {Gies}, so a cutoff of 100 modes represents an adequate computational investment. At this cutoff, for $T\approx5\textrm{nK}$, one estimates an error of $\sim10\%$ in thermal population which does not qualitatively affect our results.

Time-dependent simulations were performed using full HFB for $T=5\textrm{nK}$
for a vortex at $a=0.5$ radial harmonic oscillator units from the
axis, with a Bogoliubov mode cutoff of 100, using the time-independent initial state. In all the simulations, the vortex
precession describes a circle. There is no evidence of dissipation
during the time of the simulation of 10 trap periods. The precessional
frequency $\Omega_{LS}=\textrm{0.3794}\Omega_{r}$ was estimated using least squares, in good agreement with the value of $\Omega=\textrm{0.3727}\Omega_{r}$, as predicted in the time-independent calculations. The error in the prediction scales as the number of computational basis functions which, for practical reasons, is only 209 in these calculations. For 839 computational basis functions, the time-independent calculations predict a value of $\Omega=\textrm{0.3761}\Omega_{r}$.

\begin{figure}
\includegraphics[
  scale=0.3,
  angle=0]{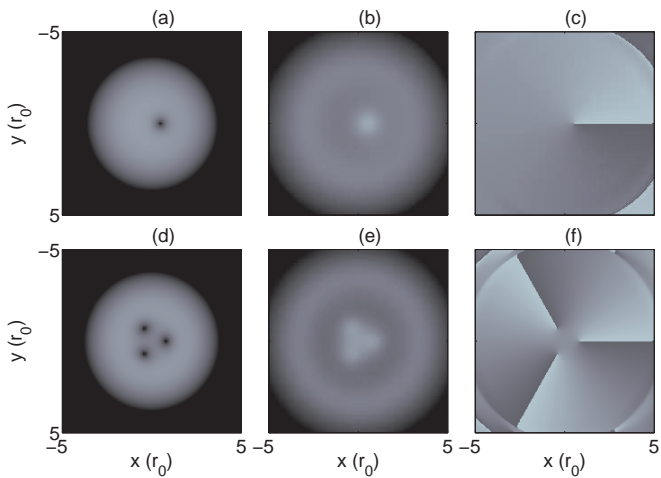}

\caption{Vortex precession in HFB time-dependent simulations showing (a) condensate density, (b) thermal density, (c) condensate phase for a single vortex at $a=0.5$, and (d) HFB condensate density, (e) thermal density, (f) condensate phase for a triangular vortex lattice for vortices at $a=1.65$. All units are given in radial harmonic oscillator units $r_{0}=\sqrt{\hbar /m\Omega_{r}}$ for distance and $E_{0}=\hbar \Omega_{r}/2$ for energies and frequencies.
\label{cap:fig3}}
\end{figure}

Figures 2(a),(b),(c) show respectively contour plots for the condensate
density, the thermal density, and the condensate phase in the x-y
plane. We note the phase discontinuity in figure 2(c) indicating a singly charged
vortex.

Time-dependent HFB simulations were also done for a triangular vortex lattice at $T=5\textrm{nK}$, with vortices symmetrically positioned at $a=1.65$ radial harmonic oscillator units from the axis, again
using time-independent calculations for the initial state. For this symmetric case, the vortex precession again describes a circle, and there is no evidence of dissipation during the time of the simulation of
10 trap periods. Figures 2(d),(e),(f) show respectively contour plots for the condensate
density, the thermal density, and the condensate phase in the x-y
plane. We note the phase discontinuity in figure 2(f) indicating three
singly charged vortices. Non-equilibrium studies of vortex lattices using the time-dependent HFB formalism will form the basis of future investigation.

In conclusion, we have performed HFB calculations for an off-axis vortex using the continuity equation to predict the precessional frequencies, which we found to be uncorrelated with the LCLS energies, in agreement with \cite{Isoshima2}. The sense and frequency of the vortex precession extrapolate smoothly from the zero-temperature GPE behaviour removing all previous anomalies. These results were confirmed using time-dependent HFB simulations.

This work was funded through the New Economy Research Fund contract NERF-UOOX0703: Quantum Technologies.

\end{document}